\documentclass[twocolumn,trackchanges]{aastex631}

\usepackage{color}

\shorttitle{The Milky Way globular cluster NGC~6355}
\shortauthors{Andr\'es E. Piatti}

\begin{document}

\title{Surviving tidal tails around the Milky Way bulge globular cluster NGC~6355}

\author[0000-0002-8679-0589]{Andr\'es E. Piatti}
\affiliation{Instituto Interdisciplinario de Ciencias B\'asicas (ICB), CONICET-UNCUYO, Padre J. Contreras 1300, M5502JMA, Mendoza, Argentina}
\affiliation{Consejo Nacional de Investigaciones Cient\'{\i}ficas y T\'ecnicas (CONICET), Godoy Cruz 2290, C1425FQB,  Buenos Aires, Argentina}

\correspondingauthor{Andr\'es E. Piatti}
\email{e-mail: andres.piatti@fcen.uncu.edu.ar}

\begin{abstract}

We present results of the analysis of a set of images obtained in the field of 
the Milky Way bulge globular cluster NGC~6355 using the Dark Energy Camera, which is 
attached to the 4m Blanco telescope of the Cerro-Tololo Interamerican Observatory. 
We dealt with a heavy differential absorption across the observed field, a
crowded field star population, and the superposition of field stars on to the
cluster color-magnitude diagram main features to produce an intrinsic cluster stars
density map. The resulting stellar density map reveals the presence of an extended 
envelope, a tidal tail, and scattered debris; the tidal tails pointing toward the
Milky Way center. Such extra-tidal overdensities, detected above the mean star field 
density, resulted to be between four and six times larger that the local star field density
fluctuation. They have also been recently generated by two independent studies which
performed numerical simulations of synthetic tidal tails of Milky Way globular clusters. 
These results contrast with previous theoretical speculations about the possibility to detect
tidal tails of globular clusters with chaotic orbits because they would be washed out
after they were generated.
\end{abstract}


\section{Introduction} 

Recently, \citet{callinghametal2022} chemo-dynamically established a group of 
42 globular clusters which belong to the Milky Way bulge. None of them is included 
in the stringent compilation of Milky Way globular clusters with studies of their
extra-tidal structures \citep{zhangetal2022}. These extra-tidal features are fundamental 
for our understanding
of their association with destroyed dwarf progenitors \citep{carballobelloetal2014,mackeyetal2019};
whether they formed in dark matter minihaloes \citep{starkmanetal2020,bv2021,wanetal2021}; 
their dynamical history as a consequence of the interaction with the Milky Way 
\citep{hb2015,deboeretal2019,pcb2020}; among others. These topics result in compelling motivations
to detect and characterize tidal tails in bulge globular clusters.

Tidal tail stars are those no longer bound to the globular cluster where they formed, 
and because of that they are located beyond the cluster Jacobi radius ($r_J$). In order
to abandon the cluster, they have increased their velocities, so that their escape velocities 
differentiate from the velocities of cluster members. For this reason, looking for
tidal tail stars from proper motions with values similar to the mean cluster proper motion
could be  misleading. Projection effects of the tidal tails, the intrinsic
kinematic agitation of the tidal tail \citep{wanetal2023}, and the Milky Way
gravitational interaction can additionally make the proper motions of tidal tail stars 
different from the mean cluster proper motion. Conversely, it is possible to find field 
stars projected along the cluster line of sight that share proper motions similar to cluster 
stars \citep{piattietal2023}. For instance, \citet{sollima2020}
claim the detection of high significance tails around NGC~7099 using 
{\it Gaia} DR2 proper motions \citep{lindegrenetal2018,gaiaetal2018b}, while 
\citet{piattietal2020} did not reproduce these using deep DECam \citep{flaugheretal2015} 
photometry. 

Two main different processes are thought to be responsible for stars to escape a globular
cluster, namely: two-body relaxation, and tidal effects caused by the 
Milky Way's gravitational field \citep[][and reference therein]{piattietal2019b}.
Hence, searches for tidal tail stars have long been focused on faint main sequence stars
(low-mass stars) because they are more numerous \citep{carballobelloetal2012}. 
However, when dealing with bulge globular clusters, and particularly with those affected by
heavy differential interstellar absorption, reaching the cluster low main sequence
could be a defying challenge for wide-field ground imaging \citep{kaderetal2023}. 
Additionally, searching tidal tails from filtering stars along the cluster color-magnitude 
diagram is constrained by the contamination of field stars located along the cluster features. 
This is because it is not possible to select a field star comparison field to decontaminate the
cluster color-magnitude diagram with the certainty {\it a priori} that it does not contain cluster
tidal tail stars. Precisely, the composite stellar population towards the Milky Way
bulge is highly dominated by field stars superimposed on to the cluster main sequences 
\citep{granetal2022}.

In this study we report the detection of tidal tails around the Milky Way bulge globular 
cluster NGC~6355, which was recently analyzed in detail by \citet{souzaetal2023} from
high-resolution spectroscopy and Hubble Space telescope imaging. NGC~6355 is a 13.2$\pm$1.1
Gyr old bulge globular cluster with a mean metallicity [Fe/H] = -1.39$\pm$0.08 dex,
placed at a distance of 8.54$\pm$0.19 kpc from the Sun. In Section 2 we present the
wide field observational data obtained to investigate the existence of tidal tails in the 
outskirts of the cluster. In Section 3 we build a cluster stellar density map, while in
Section 4 we discuss the resulting extra-tidal structures. In Section 5 we summarize the 
main conclusions of this work.

\begin{figure}
\includegraphics[width=\columnwidth]{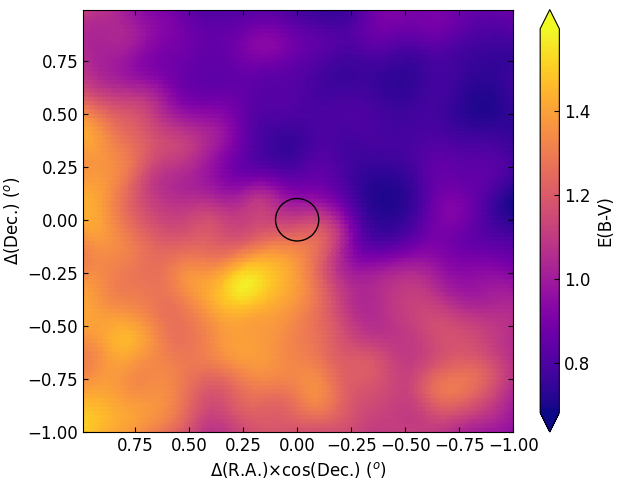}
\caption{Interstellar extinction ($E(B-V)$) map across the observed NGC~6355 field.
The black circle represents the cluster radius \citet{piattietal2019b}.}
\label{fig1}
\end{figure}

\section{Observational data}

We obtained a collection of images with the Dark Energy Camera \citep[DECam,][]{flaugheretal2015}, 
which assembles 62 identical chips with a scale of 0.263 arcsec pixel$^{-1}$, thus providing a 3 
deg$^2$ field of view. DECam is attached to the prime focus of the 4-m Blanco telescope at the 
Cerro Tololo Inter-American Observatory (CTIO). The images with a quality better than 0.6
arcsec were taken as part of the observing program 2023A-627924 (PI: A. Piatti) and consist of 
4$\times$400s $g$ and 1$\times$250s + 2$\times$400s $i$ exposures, respectively. The image
processing was fully conducted by the DECam community pipeline team, which applied the highest 
performance instrumental calibrations, and managed any peculiar image behavior. Among the
tasks carried out on the images are trimming of bad edge pixels, generating reference bias and 
dome flat files taken for the observing night to remove the instrumental signature from the 
science data, resampling the images to a standard orientation and pixel scale at a standard 
tangent point, and performing photometric calibrations.

The finally processed images were used to build a master table containing for each measured
source the equatorial coordinates, $g$ and $i$ magnitudes with their respective errors, 
sharpness, and $\chi$ values. We assigned R.A. and Dec. coordinates using the
{\sc wcstool} package\footnote{http://tdc-www.harvard.edu/wcstools/}. In order to measure
$g$ and $i$ magnitudes, we split the DECam images into nine equal squared areas of $\sim$ 0.67 
deg per side, thus performing robust quadratically spatially varying point spread function (PSF)
fitting for each subfield. In doing this, we used the {\sc daophot/allstar} suite of 
programs \citep{setal90} and two groups of PSF stars selected interactively, with $\sim$ 2200 
stars deg$^2$ and $900$ stars deg$^2$, respectively. The later contains the brightest, least 
contaminated PSF stars used to generate a preliminary PSF, which in turn was used to eliminate
neighboring stars from the former PSF group, and thus obtain a final clean fitted PSF.
We applied the resulting PSF to identified sources in each subfield using 
the {\sc allstar} program, iterating the process three times from images generated by 
subtracting known sources to measure magnitudes of fainter stars. We matched $g$ and $i$ image 
tables using the {\sc iraf.immatch@wcsmap} task, and then stand-alone {\sc daomatch/daomaster} 
routines\footnote{programs kindly provided by P.B. Stetson}. With the aim of removing bad 
pixels, unresolved double stars, cosmic rays, and background galaxies from the photometric 
catalog, we kept 888941 sources with $|$sharpness$|$ $<$ 0.5. 

\begin{figure}
\includegraphics[width=\columnwidth]{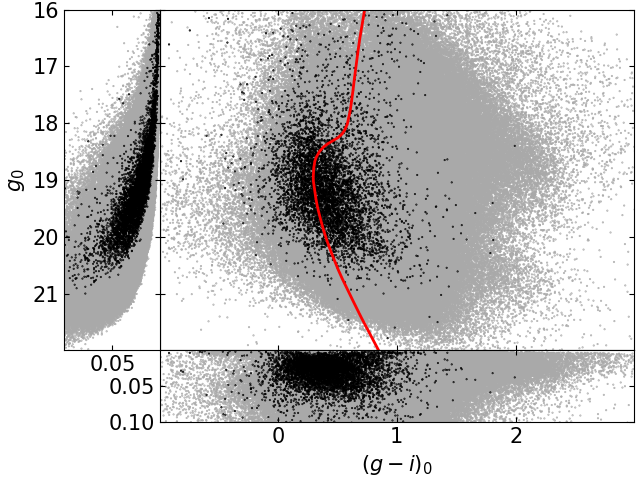}
\caption{Color-magnitude diagram for all the measured stars in the field of NGC~6355, 
with their respective photometric errors. Black and gray dots represent stars located 
inside and outside a circle of radius equals to three times the cluster's half-mass radius 
(0.022 deg), respectively. The red line is an isochrone for 
the cluster astrophysical properties \citep{souzaetal2023}.}
\label{fig2}
\end{figure}

We corrected the resulting photometry from interstellar absorption using the $E(B-V)$ values
according to the positions of the stars in the sky and using the $A_\lambda/A_V$ coefficients 
given by \citet{wch2019}. The $E(B-V)$ values were obtained by interpolation in the 
interstellar extinction map built by \citet{sf11}, provided by the NASA/IPAC Infrared Science 
Archive\footnote{https://irsa.ipac.caltech.edu/} for the entire analyzed area.
Figure~\ref{fig1} shows the relative large variation of relatively high $E(B-V)$ values
across the observe cluster region. We used the reddening corrected magnitudes and colors
to produce Figure~\ref{fig2}, which illustrates the distribution of the
measured stars in the color-magnitude diagram. Figure~\ref{fig2} highlights 
the cluster's color-magnitude diagram features. We used all the measured stars located within a 
circle of radius equal to three times the cluster's half-mass radius 
\citep[$r_h$=0.022 deg;][]{piattietal2019b} in order to minimize the field contamination. 
Figure~\ref{fig2} also shows that the photometric uncertainties of magnitude and color for the 
faintest  observed cluster's stars are smaller than $\sim$ 0.05 mag. NGC~6355 appears projected 
onto a crowded star field (gray dots in Figure~\ref{fig2}).

\begin{figure}
\includegraphics[width=\columnwidth]{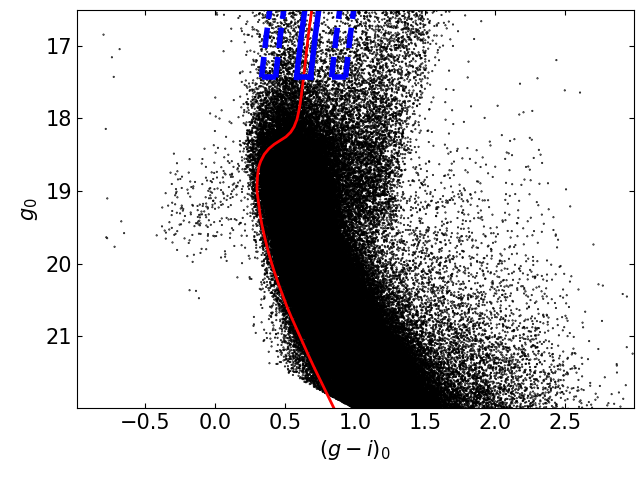}
\caption{Synthetic Milky Way color-magnitude diagram generated using the TRILEGAL code 
for a circular region centered on NGC~6355 and with a radius equals to the
cluster radius. The red line is an isochrone for the cluster astrophysical properties
\citep{souzaetal2023}, while the solid and dashed blue lines  enclose the selected region
and two control regions, respectively.}
\label{fig3}
\end{figure}

\section{Data analysis}

With the aim of disentangling whether NGC~6355 presents tidal tails, we need to construct
a stellar density map from stars sharing cluster features. Since selecting stars placed
along the cluster main sequence can incur in a misleading analysis, because of the large
amount of field stars superimposed on it, we decided to firstly explore the distribution of
Milky Way stars across the entire color-magnitude diagram. For that purpose, 
we used TRILEGAL\footnote{http://stev.oapd.inaf.it/cgi-bin/trilegal} 
\citep{girardietal2005}, a  stellar population synthesis code that  allows
changes in the star formation rate, age-metallicity relation, initial mass function, geometry 
of Milky Way components, among others, and generated a synthetic color-magnitude
diagram for a region centered on NGC~6355 with a radius equals to the cluster radius 
\citep[0.1 deg][]{piattietal2019b}. The setup of the TRILEGAL code was as follows:
1) DECam photometric system, with a limiting magnitude of $g,i$ = 23 mag,
and a distance modulus resolution of the Milky Way components of 0.1 mag. 2)
Initial mass function according to \citet{kroupa02}; a binary fraction of 0.3 with a
mass ratio larger than 0.7. 3) Interstellar extinction modeled by an 
exponential disk with a scale height $h_z$ = 0.1 kpc , a scale length $h_R$ =  3.2 kpc,
and a visual absorption variation $\partial$$A_V$/$\partial$$R$ = 0.15 mag/kpc
\citep{lietal2018a}, with the Sun position at $R_{\odot}$ = 8.3 kpc and $z_{\odot}$ = 
15 pc \citep{monteiroetal2021}. 4) Milky Way halo halo represented by an oblate $r^{1/4}$ 
spheroid with an effective radius $r_h$ =  2.7 kpc, an oblateness $q_h$ = 0.6, and
$\Omega$ = 0.0001 $M_{\odot}$/pc$^3$. The halo star formation rate and
age-metallicity relationship are those given by \citet{rn1991}. 5) Thin disk
described by an exponential disk along $z$ and $R$ with a scale height 
increasing with age: $h_z$ = 94.7$\times$(1 + age/5.5$\times$10$^9$)$^{1.67}$,
the scale length $h_R$ = 2.9 kpc, and there is an outer cutoff at 15 kpc. 
We used a two-step star formation rate, the age-metallicity relationship  with $\alpha$ enrichment
given by \citet{f1998}, and $\Sigma$ =  55.4 $M_{\odot}$/pc$^2$. 6) Thick disk
also represented by an exponential disk in both $z$ and $R$ directions, with
scale height $h_z$ = 0.8 kpc, scale length $h_R$ =  2.4 kpc, and an outer
cutoff at 15 kpc; $\Omega$ = 0.001  $M_{\odot}$/pc$^3$. We adopted a
constant star formation rate and $Z$ = 0.008 with $\sigma$[M/H] = 0.1 dex. 
7) Bulge modeled as a triaxial component with a scale length of 2.5 kpc and a truncation
scale length of 95 pc,  in addition to $y/x$ and $z/x$ axial ratios of 0.68 and 0.31, 
respectively. The angle between the direction along the bar and that from the Sun to the 
Milky Way center was set in 15 deg, and $\Omega$ = 406 $M_{\odot}$/pc$^3$. Its star formation
rate and age-metallicity relationship were taken from \citet{zoccalietal2003}.
The above values adopted for the different parameters are those obtained from the
fits of their observed behavior so that we assumed that they reliably reproduce the different 
composite stellar population characteristics \citep{vanhollebekeetal2009,girardietal2012}.

\begin{figure}
\includegraphics[width=\columnwidth]{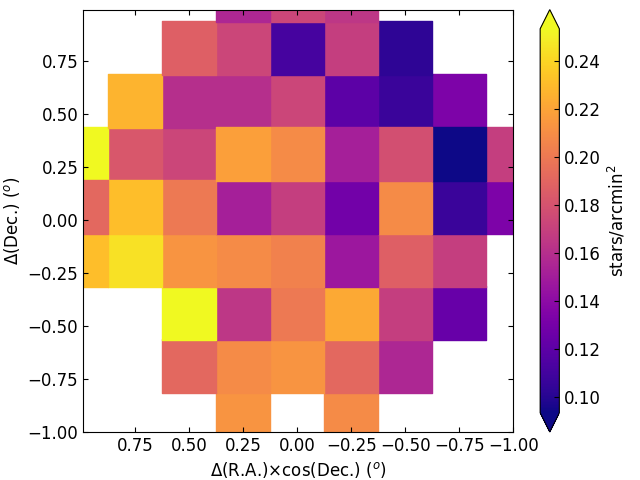}
\caption{Milky Way synthetic stellar density map built from star counts of stars 
distributed within the red giant branch segment, generated using TRILEGAL 
(see text for details).}
\label{fig4}
\end{figure}

Figure~\ref{fig3} depicts the resulting Milky Way synthetic color-magnitude diagram,
where we superimposed for comparison purposes a theoretical isochrone corresponding to the 
cluster age and metallicity. As can be seen, the red giant branch seems to be the
least cluster color-magnitude diagram contaminated region, so that we devised a segment to 
select stars 
extending from $g_0$ = 16.0 mag down to 17.5 mag and with a color width of $\Delta$($g-i$)$_0$ 
= 0.1 mag. The color width resulted in a compromise between maximizing the number of 
cluster stars and minimizing that from field stars. We also devised two control regions 
placed at both sides of the
red giant branch segment. For a more precise membership assessment within the cluster radius, we need 
additional information, such as proper motions, radial velocities,  and metallicities. 
Nevertheless, field stars placed within the blue boundaries in Figure~\ref{fig3} are 
expected to be distributed smoothly varying their densities throughout the entire field
(see Figure~\ref{fig4}), so that cluster stars will arise clearly as particular 
shaped stellar overdensities forming an extended envelope, extra-tidal debris, or tidal 
tails \citep{pcb2020}.

\begin{figure}
\includegraphics[width=\columnwidth]{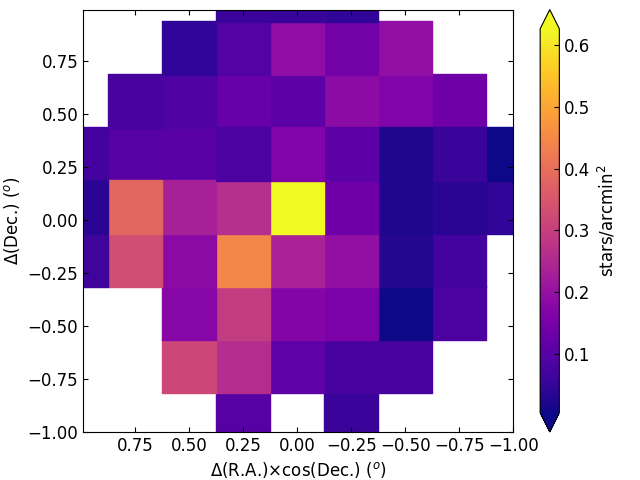}
\caption{Same as Figure~\ref{fig4} using the present DECam observations.}
\label{fig5}
\end{figure}

We also generated synthetic color-magnitude diagrams from the TRILEGAL code for 
81 adjacent squared regions of 0.0625 deg$^2$ each, which cover the entire observed DECam 
field  uniformly. From them, we counted the number of stars located within the 
red giant branch segment drawn in Figure~\ref{fig3}, and built the corresponding
stellar density map (see Figure~\ref{fig4}). Additionally, we used the TRILEGAL
squared regions and the present DECam data to build the observed stellar density map of 
Figure~\ref{fig5}. These density maps are useful tools to identify extra-tidal features 
distributed not uniformly around the cluster's main body, as it is the case of the 
presence scattered debris, tidal tails, etc. This can be accomplished by subtracting 
from the stellar densities of Figure~\ref{fig5} those of Figure~\ref{fig4}. The
resulting normalized background subtracted  residual stellar densities are shown 
in Figure~\ref{fig6}.

Besides of using the 81 TRILEGAL squared regions, we also took advantage of the whole
DECam data set to build a stellar density map of stars located within the red giant branch 
segment, employing the \texttt{scikit-learn} software machine learning library 
\citep{scikit-learn} 
and  its \texttt{Gaussian} kernel density estimator (KDE). We explored the bandwidth space
with values from 0.01$\degr$ up to 0.1$\degr$ in steps of 0.01$\degr$ and applied KDE 
for each of them using a grid of 200$\times$200 boxes onto the DECam field.
We finally adopted a bandwidth of 0.05$\degr$ as the optimal 
value, as guided by \texttt{scikit-learn}. Figure~\ref{fig7} shows the resulting 
density map, once the mean background stellar density was subtracted and those 
values normalized to the standard deviation of the background stellar density (see
Section~4 for details).

\section{Analysis and discussion}

NGC~6355 is projected on to a heavily reddened region, with $E(B-V)$ values that vary 
from 0.65 mag up to 1.6 mag across the DECam field, as is shown in Figure~\ref{fig1}. 
Figure~\ref{fig4} shows that the composite field stellar population also seems not to be 
distributed uniformly, as judged by the observed slight spatial trend in the stellar 
density of Milky Way red giant branch stars located within the devised segment of 
Figure~\ref{fig3}. The observed overall increase in the stellar density and in the
interstellar absorption from the northwestern to the southeastern regions of the DECam 
field agrees well with the direction toward the Galactic center. Curiously, the
regions more severely affected by interstellar absorption are those with more
red giant branch segment stars. Both, variable interstellar absorption and stellar 
density prevent us of using the outermost region of the DECam field as the reference one 
to clean the field contamination in the cluster color-magnitude diagram. 

\begin{figure}
\includegraphics[width=\columnwidth]{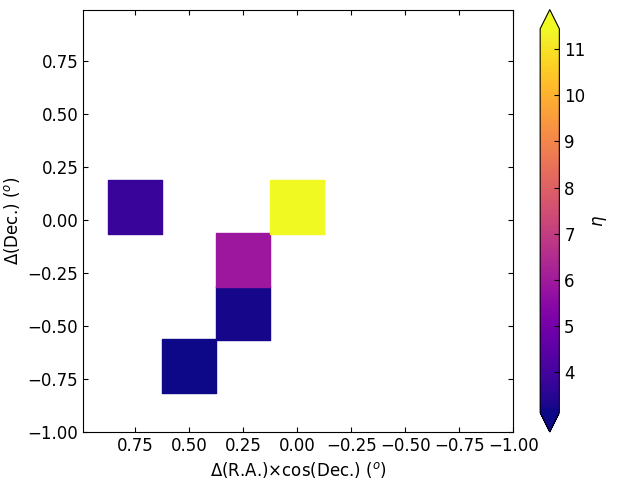}
\caption{Map of normalized background subtracted stellar densities.
 Only regions with $\eta$ $\ge$ 3 are plotted (see text for details).}
\label{fig6}
\end{figure}

We adopted a straightforward approach to clean the star counts in the cluster red giant 
branch segment from field stars across the observed DECam field (Figure~\ref{fig5}),
which consists in subtracting to the observed stellar density that measured in the Milky 
Way synthetic DECam field (Figure~\ref{fig4}), for each individual squared region.
In order to highlight possible cluster star overdensities throughout the DECam field,
we defined a significance level as the deviation from the background 
level in units of its standard deviation, that is,  $\eta$ = (signal $-$ background)/standard 
deviation. We computed the mean field stellar density and its standard deviation from 
Figure~\ref{fig4}, which resulted to be 0.18 $\pm$ 0.04 stars/arcmin$^{-2}$.
We then divided the above resulting subtraction (observed - synthetic stellar densities)
by 0.04, and kept significance level $\eta$ $>$ 3. Figure~\ref{fig6} shows that, in
addition to the cluster area at the center of the DECcam field, there are also some
regions toward the southeast from the cluster center that contain cluster
extra-tidal stars. We applied the same procedure for the two control fields
and found that none of the 81 TRILEGAL fields account for $\eta$ $>$ 3;
the mean $\eta$ value resulted to be -0.1$\pm$0.5, which means that the adopted mean stellar 
background level from Figure~\ref{fig4} is slightly larger than that observed.

\begin{figure}
\includegraphics[width=\columnwidth]{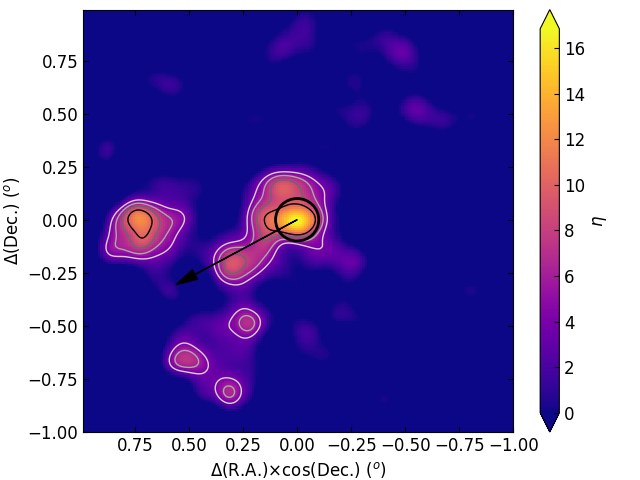}
\caption{Map of normalized background subtracted stellar densities of stars 
distributed within the red giant branch segment 
(see Figure~\ref{fig3}). Contours for $\eta$ = 4, 6, 8, and 10 are drawn with with to black lines, 
respectively. The arrows indicates the direction to the Milky
Way center. The black circle represents the cluster radius 
\citep[0.1 deg;][]{piattietal2019b}, which is equivalent to 14.9 pc.}
\label{fig7}
\end{figure}

Such extra-tidal structures are clearly seen in Figure~\ref{fig7},
where we superimposed contour levels corresponding to $\eta$= 4, 6, 8, and 10 using
gray-colored lines, so that the darken the line the higher the stellar density. 
The cluster radius and the direction toward the Galactic center are also
indicated with a black circle and an arrow, respectively. 
Inside the cluster radius, we found overdensities above the mean
stellar field density larger than 16 times its standard deviation. As can be seen, NGC~6355
presents an extra-tidal extended envelope azimuthally not symmetric, a tidal tail, and 
scattered stellar debris alongside the tidal tail. As far as we are aware, this is the
first Milky Way budge globular cluster with detected extra-tidal structures.

The outcome is somehow surprising according to the theoretically expected scenario.
\citet{mestreetal2020} compared the behavior of simulated streams embedded 
in chaotic and regular regions of the phase space and found that typical 
gravitational potentials of host galaxies can sustain chaotic orbits, which in turn 
do reduce the time interval during which streams can be detected. This explains
why tidal tails in some globular clusters are washed out after they are generated 
to the point at which it is impossible to detect them. NGC~5139, with a Galactocentric 
distance of 6.5 kpc, has been until now the innermost globular cluster 
with observed tidal tails \citep{pcb2020}; NGC~63655 is at 0.93 kpc from the Galactic center \citep{bv2021}. 
According to the cluster's angular momentum along the direction perpendicular to the 
Galactic plane \citep{souzaetal2023}, the mass of the Milky Way bulge \citep{valentietal0216}, 
and the regime of chaotic orbits given by \citet{cp2003}, NGC~6355 can be added
to the group of objects with chaotic orbits.

At the present time, it is still a topic of debate the presence of tidal tails in some 
globular clusters and  the absence of them in others. \citet{pcb2020} explored
different structural and kinematical parameter spaces, and found that
globular clusters behave similarly, irrespective of the presence of tidal tails or
extended envelopes, or the absence thereof. \citet{zhangetal2022} showed that globular
clusters with extended envelopes or tidal tails have apogalactocentric distances
$\ga$ 5 kpc, a behavior previously noticed by \citet{piatti2021c}, who suggested that
the lack of detection of tidal tails in bulge globular clusters could be due to the reduced 
diffusion time of tidal tails by the kinematically chaotic nature of the orbits of these 
globular clusters \citep{kunduetal2019b}, thus shortening the time interval during which 
the tidal tails can be detected. Recently, \citet{weatherfordetal2023} reexamined
the behavior of potential escapers in globular clusters dynamically evolving along
chaotic orbits and found diffusion times shorter than 100 Myr. If this were the case
for NGC~6355, the observed tidal tails would reveal an ongoing disruption mass process.
The tidal tails in NGC~6355 points to the direction toward the Galactic center, which 
coincides with the theoretical predictions for the innermost parts of tidal tails 
\citep{montuorietal2007}.

\begin{figure}
\includegraphics[width=\columnwidth]{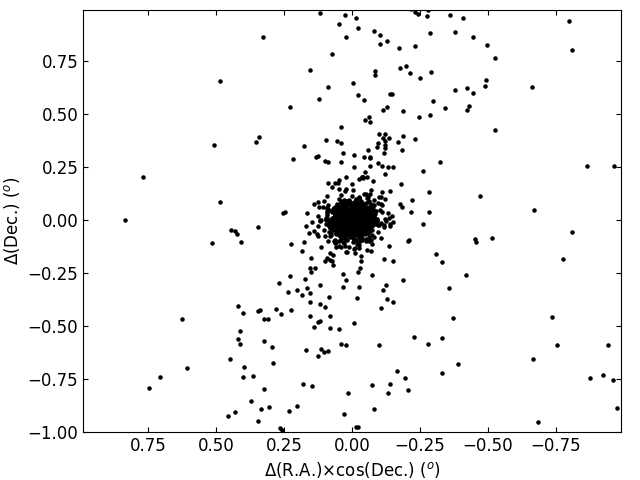}
\caption{Spatial distribution of stars with action ratios
within 1$\sigma$ the cluster mean values. Data points arise from the 
\citet{grondinetal2024}'s models.}
\label{fig8}
\end{figure}

Recently, \citet{grondinetal2024} generated a catalog of globular cluster extra-tidal
mock stars from simulations of three-body dynamical encounters in globular cluster cores.
The catalog provides celestial coordinates, proper motions, radial velocities, 
the actions $J_R$, $J_z$, and $J_{\phi}$, among other stellar properties for each simulated
star. These features were obtained by integrating stellar orbits in seven Milky Way
scenarios where static and time-varying gravitational potentials, the structure
of the disk, the shape of the dark matter halo, and the perturbations of the Large Magellanic
Cloud are considered. They found that the relations between different actions are similar
for cluster stars and for those stars belonging to their tidal tails. Particularly,
the plane $J_{\phi}$/$J_{tot}$ versus ($J_z - J_R$)/$J_{tot}$, where $J_{tot}$ = $J_R$ +
$|$$J_{\phi}$$|$ + $J_z$, resulted to be the most suitable for identifying cluster stars
distributed in their tidal tails. We computed the mean and dispersion of both ratios
for  mock stars generated by \citet{grondinetal2024} located within a radius 
equals to the NGC~6255 radius, and then filtered
all the mock stars located within the DECam field. Figure~\ref{fig8} illustrates the
resulting spatial distribution of cluster and extra-tidal mock stars with action ratios
within 1$\sigma$ the cluster mean values. As can be seen, the presence of tidal tails
is revealed, with that toward the southeast resembling the observed one (see Figure~\ref{fig7}).
Nevertheless, when comparing Figures~\ref{fig7} and \ref{fig8} some difference
remains, namely: the lack of an observed northwestern tail, and the appearance of 
an scattered observed extra-tidal debris located Eastward the cluster.

Models of the formation and evolution of extra-tidal features around Milky Way globular clusters
have also been computed by \citet{ferroneetal2023}. They used a set of codes called Globular
Clusters' Tidal Tails (GCsTT) to simulate streams that vary with the Milky Way potential by
making use of three different profiles, either containing a central spheroid, not containing 
one, or containing a stellar bar. \citet{ferroneetal2023} analyzed 159 globular clusters,
for which 6D phase-space information, masses, and sizes are available, and classified the
simulated extra-tidal structures as main tidal structures if the fraction of stripped stars
from the globular cluster is larger than 10$\%$. They analyzed 40 out of the 42 bulge globular
clusters classified by \citet{callinghametal2022}, and 36 of them present main tidal structures,
including NGC~6355. This implies that tidal tails among bulge globular clusters seem to be
a more common phenomenon that previously thought. In this sense, NGC~6355 arises as the
first bulge globular clusters for which its outermost regions are studied with the aim of
identifying extra-tidal features. The outcomes by \citet{ferroneetal2023} point to the need
of observational campaigns and data analysis strategies of Milky Way globular clusters
to explore their outskirts. Such data will help confirming and improving the current 
theoretical inputs of numerical simulations, and hence also our knowledge of the formation and
evolution of the Milky Way.

\section{Conclusions}

We analyzed wide-field DECam $g,i$ imaging around the Milky Way bulge globular cluster
NGC~6355, with the aim of searching for extra-tidal structures. The cluster was observed
as part of our CTIO 4-m Blanco telescope 2023A-627924 observing program, thus obtaining
relatively deep photometry of the cluster's outskirts. Because of the heavy differential
reddening and crowded star field population along the cluster line of sight, and the 
superposition of field stars on to the cluster color-magnitude diagram, we devised an analysis 
strategy that minimizes those effects. Interstellar extinction was properly corrected using 
reddening values retrieved from the NASA/IPAC Infrared Science Archive, while the presence of 
field stars in the cluster color-magnitude diagram was constrained by selecting the least 
contaminated region at the upper observed part of the cluster red giant branch. Such a
red giant branch segment was chosen from the inspection of synthetic color-magnitude
diagrams generated using the TRILEGAL code.

The stellar density map built using the synthetic star field density spatial distribution
to decontaminate the observed one reveals the presence of extra-tidal overdensities
above the mean stellar field density, which are between 4 and 6 times larger than the star field 
density dispersion. The extra-tidal
stellar overdensities comprise those of an extended envelope, a tidal tail, and scattered
debris; the tidal tail pointing to the direction toward the Milky Way center. 
We compared our results with two recent independent catalogs of simulated tidal tails
around globular clusters, and confirmed the present identification of tidal tails around
NGC~6355; although previous theoretical speculations discouraged the possibility to
detect them because they were meant to be washed out after they were generated.

The advance produced in the generation of synthetic tidal tail features of any Milky Way 
globular cluster urge us to carry out observational surveys of the outskirts of 
globular clusters, particularly those still not studied, with the aim of providing a 
robust observational counterpart  for the simulations, which will result in a 
feedback to the physics considered in the computational generation of tidal tails. In this 
work, we provide a first example of such an endeavor, confirming that the observed tidal 
tails of NGC~6355 are also predicted by numerical simulations.

\begin{acknowledgements}
We thank the referee for the thorough reading of the manuscript and
timely suggestions to improve it. 

This project used data obtained with the Dark Energy Camera (DECam), which was constructed by 
the Dark Energy Survey (DES) collaboration. Funding for the DES Projects has been provided by the 
US Department of Energy, the US National Science Foundation, the Ministry of Science and Education 
of Spain, the Science and Technology Facilities Council of the United Kingdom, the Higher 
Education Funding Council for England, the National Center for Supercomputing Applications at the
University of Illinois at Urbana-Champaign, the Kavli Institute for Cosmological Physics at the
University of Chicago, Center for Cosmology and Astro-Particle Physics at the Ohio State 
University, the Mitchell Institute for Fundamental Physics and Astronomy at Texas A\&M University, Financiadora de Estudos e Projetos, Funda\c{c}\~{a}o Carlos Chagas Filho de Amparo \`a Pesquisa 
do Estado do Rio de Janeiro, Conselho Nacional de Desenvolvimento Cient\'{\i}fico e Tecnol\'ogico 
and the Minist\'erio da Ci\^{e}ncia, Tecnologia e Inova\c{c}\~{a}o, the Deutsche Forschungsgemeinschaft and the Collaborating Institutions in the Dark Energy Survey.

The Collaborating Institutions are Argonne National Laboratory, the University of California at 
Santa Cruz, the University of Cambridge, Centro de Investigaciones En\'ergeticas, Medioambientales 
y Tecnol\'ogicas–Madrid, the University of Chicago, University College London, the DES-Brazil Consortium, the University of Edinburgh, the Eidgenössische Technische Hochschule (ETH) Zürich, 
Fermi National Accelerator Laboratory, the University of Illinois at Urbana-Champaign, the 
Institut de Ci\`encies de l’Espai (IEEC/CSIC), the Institut de F\'{\i}sica d’Altes Energies, 
Lawrence Berkeley National Laboratory, the Ludwig-Maximilians Universit\"at München and the
associated Excellence Cluster Universe, the University of Michigan, NSF’s NOIRLab, the University 
of Nottingham, the Ohio State University, the OzDES Membership Consortium, the University of Pennsylvania, the University of Portsmouth, SLAC National Accelerator Laboratory, Stanford University, the University of Sussex, and Texas A\&M University.

Based on observations at Cerro Tololo Inter-American Observatory, NSF’s NOIRLab (NOIRLab Prop. ID 
2023A-627924; PI: A. Piatti), which is managed by the Association of Universities for Research in Astronomy (AURA) under a cooperative agreement with the National Science Foundation.

Data for reproducing the figures and analysis in this work will be available upon request
to the author.

\end{acknowledgements}


\end{document}